\documentclass[12pt]{article}
\input epsf.sty
\newcommand{\ra}{\rangle}
\newcommand{\la}{\langle}

\newcommand{\OO}{{\cal O}}

\newcommand{\wt}{\widetilde}

\newcommand{\ta}{c_1|0\ra}
\newcommand{\be}{\begin{equation}}
\newcommand{\ee}{\end{equation}}
\newcommand{\ben}{\begin{eqnarray}\displaystyle}
\newcommand{\een}{\end{eqnarray}}
\newcommand{\refb}[1]{(\ref{#1})}
\newcommand{\p}{\partial}
\newcommand{\sectiono}[1]{\section{#1}\setcounter{equation}{0}}

\begin{document}
{}~
\hfill\vbox{\hbox{hep-th/0201159}\hbox{CTP-MIT-3231}
\hbox{CGPG-02/1-2}
\hbox{PUPT-2021}
}\break

\vskip .6cm

\centerline{\large \bf
Patterns in Open String Field Theory Solutions
}

\vspace*{4.0ex}

\centerline{\large \rm Davide Gaiotto$^a$, Leonardo Rastelli$^a$,
Ashoke Sen$^b$ and Barton Zwiebach$^c$}

\vspace*{4.0ex}

\centerline{\large \it ~$^a$Department of Physics }

\centerline{\large \it Princeton University, Princeton, NJ 08544,
USA}

\centerline{E-mail: dgaiotto@princeton.edu,
        rastelli@feynman.princeton.edu}

\vspace*{1.8ex}

\centerline{\large \it ~$^b$Harish-Chandra Research
Institute}

\centerline{\large \it  Chhatnag Road, Jhusi,
Allahabad 211019, INDIA}

\centerline {and}
\centerline{\large \it Department of Physics, Penn State University}

\centerline{\large \it University Park,
PA 16802, USA
}

\centerline{E-mail: asen@thwgs.cern.ch, sen@mri.ernet.in}

\vspace*{1.8ex}

\centerline{\large \it $^c$Center for Theoretical Physics}

\centerline{\large \it
Massachussetts Institute of Technology,}

\centerline{\large \it Cambridge,
MA 02139, USA}

\centerline{E-mail: zwiebach@mitlns.mit.edu}

\vspace*{4.0ex}

\centerline{\bf Abstract}
\medskip

In open string field theory the kinetic operator
mixes matter and ghost sectors, and thus the ghost
structure of classical solutions is not universal. 
Nevertheless, we have found from numerical analysis
that certain ratios of expectation values for states involving
pure ghost excitations appear to be
universal. We give an analytic expression for these ratios 
and find good evidence that they are common to all known
solutions of open string field theory, including the tachyon
vacuum solution, lump solutions and string fields representing 
marginal deformations.  We also draw attention
to a close correspondence between the
expectation values for the pure
matter components in the tachyon vacuum solution
and those  in the solution of a simpler equation
for a  ghost number zero string field.
Finally we observe that the action of $L_0$ on the tachyon condensate 
gives a state that is approximately factorized into a matter and a ghost 
part.  

\vfill \eject

\baselineskip=16pt

\tableofcontents

\sectiono{Introduction and summary}\label{s0}

During the last few years many different kinds of solutions of open cubic
string field theory \cite{OSFT} (OSFT) have been
studied numerically. These
include the tachyon vacuum solution \cite{VAC}, tachyon lump
solutions describing various D-branes \cite{0005036,LUMP},
and
solutions representing marginal deformations \cite{MARG}.
So far, however, there has
been little progress in obtaining the analytic  
form of any of these solutions
(see \cite{attempts} for some attempts).
This is the problem that  motivates the work presented in this paper.

Analytic solutions
are much easier to obtain in vacuum string field theory
(VSFT), a version of
open
string field theory proposed to describe physics around the tachyon vacuum
(see \cite{0111129} and references cited therein, 
and \cite{recent,0112231,0201095,0201015} for some 
recent developments). 
Nevertheless, there are several
reasons that analytic solutions to OSFT are desirable. As presently
formulated VSFT requires an extremely simple but singular kinetic term.
An OSFT solution
for the tachyon vacuum would most likely allow a derivation of VSFT, and
may show if there is a simple regular form of the theory. In doing so the
solution  would allow 
a clear and complete proof of the tachyon conjectures
\cite{conj}. Moreover, solutions of OSFT
are likely to teach us interesting and useful facts about the 
open string star algebra
and its interplay with the BRST and/or Virasoro operators.
While solutions of VSFT have recently
taught us a lot about star algebra projectors, more tools
seem necessary to write OSFT solutions.

In OSFT the string field is represented by a ghost number one state in the
state  space of the combined matter-ghost conformal field theory.  Since
physics around different classical solutions (other than possibly the
vacuum solution) are described by different matter conformal field
theories with a  common ghost system, one might have naively expected that
these different classical solutions will have a common ghost structure,
and will differ from each other in their matter part.
While this is the case in VSFT, and is 
a direct consequence of its simplicity,
it is not the case in OSFT as can be easily seen by examining the
various numerical solutions obtained so far. One could hardly have expected
this structure since the kinetic operator of OSFT mixes non-trivially
matter and ghost sectors. Nevertheless one
could ask if some part of the ghost structure
is universal, namely, independent
of which classical solution we are considering.
This is one of the questions we shall address.  

We show in section \ref{s2} that
if the coefficient of any state in the string field theory solution has
such universal structure, then it is easy to find the analytic expression
for this coefficient by examining the solution describing small marginal
deformations. Marginal deformations are associated with a dimension one
boundary operator in the matter conformal field theory. To first order in
the deformation parameter the solution representing such deformation is
represented by only one state, obtained by the product of this dimension one
matter primary and the ghost ground state of dimension minus one. To second
order in the deformation parameter various other states are excited, but
the coefficients of each of these states can easily be found by explicitly
solving the string field equations to this order. This gives a way to
compute the {\it ratios} of the coefficients of various
states in this solution
to this order. If any of these ratios is universal, then this computation
determines this universal number.

While this tells us how to compute a universal ratio,
it does not tell us which ratios  
are likely to be
universal. For this we rely on
numerical results. In section \ref{s3} we take various solutions (vacuum
solution, lump
solutions and solutions representing marginal deformations) and compare the
coefficients appearing in these solutions to determine which of these
coefficients are universal. Our analysis indicates that the ratios of the
coefficients of the state $c_{-n} b_{-m} c_1|0\rangle$ to that of
$c_1|0\rangle$ may be
universal for any pair of {\it odd} integers $m$, $n$.
The explicit analytical expression for this ratio can then be computed
using the method outlined above and is given by $n \wt N^{11}_{mn} /
(n+m-1)$, where $\wt N_{mn}$ are the ghost Neumann coefficients introduced
in ref.\cite{gj2}. We are thus 
led to believe that
the string field $|\Psi\rangle$ 
representing a classical solution of OSFT contains a piece
\be
\label{mainequation}
|\Psi \rangle =
\alpha_0\, \Bigl( \,1 + \sum_{n,m = odd}{n \wt
N^{11}_{mn} \over
(m+n-1) }c_{-n} b_{-m}  \Bigr)
c_1 |0\rangle + \cdots \,.
\ee
While an infinite series of ratios are fixed, the overall constant
$\alpha_0$ is solution dependent. An immediate consistency check is
possible. It was shown in \cite{0009105,0010190} that a solution of
the string field equations in the Siegel gauge must be singlets
of an $SU(1,1)$ symmetry acting on the ghost sector.
Since, $\wt N^{11}_{mn} = \wt N^{11}_{nm}$ for the cubic vertex of OSFT,
it is clear from the above expression that this sector of the string
field is built from linear combinations of the form
$(n c_{-n} b_{-m} + m c_{-m} b_{-n})$ and these are indeed $SU(1,1)$
singlets \cite{0009105,0010190}. Of course, the prediction in
\refb{mainequation}
goes far beyond $SU(1,1)$ in that it prescribes specific linear combinations
of those singlets.

\medskip
Given that the string field equations are non-linear, the equations of
motion
of the proposed universal coefficients will receive contribution from the
non-universal coefficients. Thus the universality of the type proposed
here may seem
highly unnatural. However, it will be natural if these
coefficients were determined by a set of linear equations satisfied by the
string field. Although at present we do not know of any way to derive all
the universal coefficients this way, we show in section \ref{sposs} that
there are some linear equations which any solution of the string field
theory equations of motion is expected to satisfy. Some particular
relations involving the universal coefficients follow from these linear
equations.

\medskip  
Sections \ref{sfactor} and \ref{ssurp} are devoted 
to the study of some other aspects of the solutions of OSFT equations of 
motion.
In vacuum string field theory solutions of classical equations of motion
are factorized into a product of a matter part and a ghost part. This is
not the case in OSFT since the kinetic operator $Q_B$ involves both matter
and ghost parts. Nevertheless we show in section \ref{sfactor} that
the numerical results for the solution 
$|\Psi_0\ra$  
representing the 
tachyon
vacuum has the property that $L_0|\Psi_0\ra$ is approximately factorized
into a product of a matter part and a ghost part.  

In section \ref{ssurp} we
draw attention to a surprising feature of the
numerical results for the pure matter excitations
in the tachyon condensate $|\Psi_0\ra$ representing the vacuum solution.
These excitations are given by the action of the matter
Virasoro generators on the
ghost number one ground state $c_1|0\rangle$. Denoting by $\OO_m$ any
particular combination of matter Virasoro generators, we find that
the ratio of the coefficient of $\OO_m c_1|0\ra$ to that of
$c_1|0\rangle$ in $|\Psi_0\ra$ is very close to the ratio of $\OO_m|0\ra$
to $|0\ra$ in the solution of another equation:
\be \label{e01}
(L_0-1)|\Phi\ra + |\Phi * \Phi\ra =0 \, ,
\ee
where $|\Phi\ra$ is a ghost number zero string field. The correspondence
is tested using numerical solutions to the field equations
to level (10,30) in appendix \ref{sa}.

\sectiono{Computation of the universal coefficients} \label{s2}

We consider formulating string field theory in a background represented by
a matter boundary conformal field theory with an exactly marginal
deformation. Associated with this deformation is a dimension one matter
primary field $V$. Let $|V\rangle_m$ denote the corresponding state in the
Hilbert space of the matter CFT. Then to first order in the deformation
parameter $\lambda$, the solution of the OSFT field equations representing
this deformation is given by:
\be \label{e1}
|\Psi\ra = \lambda c_1 |0\ra_g \otimes |V\ra_m \equiv \lambda
|\chi^{(1)}\ra\, ,
\ee
where  
$|0\ra_g$ denote the SL(2,R) invariant ghost vacuum. $|\Psi\ra$ given
above satisfies the
OSFT field equation:
\be \label{e2}
Q_B|\Psi\ra + |\Psi *\Psi\ra = 0\, ,
\ee
to order $\lambda$. Let us now denote the solution to the OSFT field
equation to second order in $\lambda$ by
\be \label{e3}
|\Psi\ra = \lambda
|\chi^{(1)}\ra + \lambda^2
|\chi^{(2)}\ra\, .
\ee
Substituting this into eq.\refb{e2} and collecting terms to second order
in $\lambda$ we get,
\be \label{e4}
Q_B|\chi^{(2)}\ra = - |\chi^{(1)} * \chi^{(1)}\ra \, .
\ee
If we take $|\Psi\ra$ to be in the Siegel gauge,
\be \label{e5a}
b_0 |\Psi\ra = 0\, ,
\ee
then, by applying $b_0$ on both sides of the equation we get:
\be \label{e5}
L_0 |\chi^{(2)}\ra = - b_0  |\chi^{(1)} * \chi^{(1)}\ra \, .
\ee
We can now easily solve for $\chi^{(2)}$ by expressing it as an arbitrary
linear
combination of ghost number one states and comparing the two sides of the
above equation. One particular consistency condition required for this
equation to have a solution is that $ b_0  |\chi^{(1)} * \chi^{(1)}\ra$
should not contain any state of vanishing $L_0$ eigenvalue. This in turn
is related to the condition that $V$ is an exactly marginal operator in
the matter CFT.

\medskip
The right hand side of eq.\refb{e5} can be evaluated as follows. We write:
\be \label{e6}
b_0^{(3)} |\chi^{(1)} * \chi^{(1)}\ra_{(3)}
= (|V\ra_m *_m |V\ra_m)_{(3)} \otimes  ~_{(1)}\la 0|c^{(1)}_{-1} \,
~_{(2)}\la
0|c^{(2)}_{-1} \,\,\, b_0^{(3)} | V_{123}\ra \, ,
\ee
where $|V_{123}\ra$ is the ghost vertex, given by \cite{gj2}:
\be \label{e7}   
|V_{123}\ra = \exp\bigg(\sum_{r,s=1}^3 \sum_{{m\geq 0\atop n\ge 1}} \wt
N^{rs}_{mn} n
b^{(r)}_{-m} c^{(s)}_{-n}\bigg) c_0^{(1)} c_1^{(1)}|0\ra_{(1)}
\otimes  c_0^{(2)} c_1^{(2)}|0\ra_{(2)} \otimes  c_0^{(3)}
c_1^{(3)}|0\ra_{(3)}\, .
\ee
Here the $\wt N^{rs}_{mn}$ are Neumann
coefficients. They have cyclic symmetry $(r,s)\to
(r+1, s+1)$ with $r+3\equiv r$, $s+3\equiv s$.
In the matter sector the star product gives
\be \label{e8}
|V\ra_m *_m |V\ra_m = \alpha |0\ra_m + \cdots\, ,
\ee
where $\alpha$ is some constant, and $\cdots$ denote 
excited states in the matter sector.
The effect of the ghost vacua $\langle 0 | c_{-1}$ in the right
hand side of \refb{e6} is to delete from the ghost part of the
vertex all reference to oscillators
in the first and second state spaces.
Thus
\be \label{e66}
b_0^{(3)} |\chi^{(1)} * \chi^{(1)}\ra_{(3)}
= (\alpha |0\ra_m + \cdots)_{(3)} \otimes  ~
\exp\bigg( \sum_{m,n\ge 1} \wt N^{11}_{mn} n
b^{(3)}_{-m} c^{(3)}_{-n}\Big)
c_1^{(3)}|0\ra_{(3)} \, .
\ee
Using eqs.\refb{e5},\refb{e66},
and the
equation:
\be \label{e9}
L_0 c_{-n} b_{-m} c_1|0\ra = (m+n-1)  c_{-n} b_{-m} c_1|0\ra\, ,
\ee
we find the coefficient $\alpha_{n,m}$ of the state $c_{-n} b_{-m} 
c_1|0\ra$ in
$|\chi^{(2)}\ra$ to be:
\be \label{e10}
\alpha_{n,m} =  {n\over m+n-1} \, \wt N^{11}_{mn} \, \alpha\, .
\ee
On the other hand, the coefficient of the $c_1|0\ra$ term, obtained by
keeping the zeroth order term in the expansion of the exponential in
eq.\refb{e7}, is
\be \label{e11}
\alpha_0 = \alpha\, .
\ee
Thus we get the ratio of these terms to be 
\be \label{e12}  
r_{n,m}\equiv {\alpha_{n,m}\over \alpha_0} = {n\over m+n-1} \, \wt
N^{11}_{mn} \equiv \bar r_{n,m}\, .
\ee

Although the ratio $r_{n,m}$ has been computed for a solution of OSFT representing
a marginal deformation with  small  deformation parameter,
if this ratio is universal then $\bar r_{n,m}$ as defined in\refb{e12}
will represent the ratio
of the coefficients of $c_{-n} b_{-m} c_1|0\ra$ and $c_1|0\ra$ in any
solution of the string field theory equations of motion. In the next
section we shall examine the  
numerical results and show that these ratios do appear to be universal
provided $m$ and $n$ are  {\it odd} integers.

\sectiono{Test of universality of the coefficient of $c_{-n} b_{-m}
c_1|0\ra$} \label{s3}

In this section we present numerical evidence that the ratio $r_{m,n}$ of
the coefficients of $c_{-n} b_{-m} c_1|0\ra$ and $c_1|0\ra$ is universal,
independent of which solution we analyze. We do this by evaluating these
coefficients for various solutions obtained by using level truncation, 
and showing that in each case the
result  comes close to the prediction \refb{e12}.

\medskip
For the explicit prediction we need the  Neumann coefficients $\wt
N^{11}_{mn}$. We have:
\ben
\label{thematrix}
\wt N^{11}_{mn} &=&  {2\over 3}
{(-1)^{m+1} \over n^2 - m^2}
\Bigl (n A_m
B_n - m
A_n
B_m
\Bigr) \,, \quad m+n = ~\hbox{even},\,\,\,  m\not= n\,,  \nonumber \\
\wt N^{11}_{mn} &=& \,\,\,\,0\,, \qquad\qquad\qquad\qquad\quad
\qquad \qquad \quad m+n =
~\hbox{odd}\,, \\
\wt N^{11}_{nn} &=& {1\over 3 n} (-1)^n \,\Bigl( 2 S(n) - 1 -(-1)^n A_n^2
- 2 A_n B_n
\Bigr) \,,
\qquad S(n) = \sum_{k=0}^n  (-1)^k A_k^2 \,.\nonumber
\een
In the above the coefficients $A$ and $B$ are defined as
\be \label{ea1}
\bigg({1 + i x \over 1 - i x}\bigg)^{1/3} = \sum_{n\, even} A_n
x^n
+ i \sum_{n\, odd} A_n x^n\, , \quad
\bigg({1 + i x \over 1 - i x}\bigg)^{2/3} = \sum_{n\, even} B_n
x^n
+ i \sum_{n\, odd} B_n x^n\, .
\ee

The ghost sector of OSFT
in the Siegel gauge is
invariant under an $SU(1,1)$ symmetry having a $Z_4$ subgroup
\cite{0009105,0010190}. This is, in fact, reflected in the
symmetry  $\wt N^{rs}_{mn} = \wt N^{sr}_{nm}$, 
which implies
that $\wt N^{11}_{mn}$ is symmetric, as manifest in \refb{thematrix}.
For a string field built as a bilinear in ghosts and antighosts
acting on $c_1|0\rangle$, the condition of
$SU(1,1)$ invariance reduces to a $Z_4$ invariance \cite{0009105}:
$b_{-n} \to -n c_{-n}$, $c_{-n}  \to \frac{1}{n} b_{-n}$. As noted
in the introduction below \refb{mainequation}, the proposed
universal part of the 
string field is built as linear superpositions of terms of the
form $(n c_{-n} b_{-m} + m c_{-m} b_{-n})$ acting on $c_1|0\rangle$,
and these are readily seen to be singlets.  Since $SU(1,1)$ is a symmetry
even after
level truncation it need not be tested further and it will
suffice for us to test the universality of 
the ratios $r_{n,m}$
for $n\ge m$.

To facilitate comparison with numerical results, we
now list the predicted values of $r_{n,m}$
for various values of $m$, $n$
up to $m+n\le 10$.
Using
\refb{e12} and
the above expressions for the Neumann coefficients, we find
\ben \label{e21}  
&& \bar r_{1,1} = {11\over 27} \simeq 0.407407, \qquad \bar r_{3,1} = 
-{80\over 
729}
\simeq
-0.109739, \nonumber \\
&& \bar r_{5,1} = {1136\over 19683} \simeq 0.0577148, \qquad
\bar r_{3,3}={2099\over 98415} \simeq 0.021328\, , \nonumber \\
&& \bar r_{7,1} = -{6640\over 177147} \simeq -0.037483, \qquad \bar 
r_{5,3} =
-{17840\over 1240029} \simeq -0.0143868, \nonumber \\
&& \bar r_{9,1} = {388336\over
14348907} \simeq 0.0270638, \qquad \bar r_{7,3} = {455728 \over
43046721} \simeq 0.0105868,
\nonumber \\
&& \bar r_{5,5} =
{94979\over 14348907} \simeq 0.00661925\, .
\een
In order to show that $r_{n,m}$ for even $m,n$ are not universal in
general, we also give the value of $\bar r_{2,2}$ computed according to
eq.\refb{e12}. It is
\be \label{er22}
\bar r_{2,2} = -{19\over 729} \simeq -0.0260631\, .
\ee

\subsection{Tachyon vacuum solution}

For the tachyon vacuum solution we present the results for the ratio
$r_{m,n}$ at level $(L,3L)$ approximation for various values of $L$, and
also extrapolate the results to $L=\infty$ using a linear fit of the form:
$f(L) = a + b / L$ with constants $a$, $b$. The results are shown in
tables
\ref{t1}, \ref{t2}, and are clearly in good agreement with the predicted
values
\refb{e21}. Indeed in table \ref{t1}   
the projections differ by about 1\%
or less from the predictions.  This is also the case for
the first two columns in table \ref{t2}.
Even for the cases
where there is just one data point the values are surprisingly
close to the predictions.

In order to demonstrate that the coefficient $r_{2,2}$ does not follow the
prediction \refb{er22}, we now quote the level (10,30)
result for this
ratio. It is:
\be \label{er221020}
r_{2,2} = -0.0640389\, .
\ee
This is quite far from the prediction \refb{er22}.

\begin{table}
\begin{center}\def\st{\vrule height 3ex width 0ex}
\begin{tabular}{|l|l|l|l|l|} \hline
$L$ & $r_{1,1}$ & $r_{3,1}$ & $ r_{5,1}$  &
$r_{3,3}$
\st\\[1ex]
\hline
\hline
4 & 0.375042 &  -0.102499 &  &
\st\\[1ex]
\hline
6 & 0.386571  & -0.104743 & 0.0547758  & 0.0208371
\st\\[1ex]
\hline
8 & 0.393062 & -0.105966 & 0.05544 & 0.0209129
\st\\[1ex]
\hline
10 & 0.397214 & -0.106707 & 0.0558499 & 0.0209927
\st\\[1ex]
\hline \hline
$\infty$ &0.411545 &-0.109469  &0.0574564  &0.0212121
\st\\[1ex]
\hline \hline
$conj$ &0.407407 &-0.109739  &0.0577148  &0.021328
\st\\[1ex]
\hline
\end{tabular}
\end{center}
\caption{The numerical results for the coefficients $r_{n,m}$ for the
tachyon vacuum solution at level $(L,3L)$ approximation,
their extrapolation to $L=\infty$ via a fit $a + b/L$,
and the conjectured values.} \label{t1} \end{table}

\begin{table}
\begin{center}\def\st{\vrule height 3ex width 0ex}
\begin{tabular}{|l|l|l|l|l|l|} \hline
$L$ & $r_{7,1}$ & $r_{5,3}$ & $ r_{9,1}$  &
$r_{7,3}$ & $r_{5,5}$
\st\\[1ex]
\hline
\hline
8 & -0.0358037  & -0.014195 &  &  &
\st\\[1ex]
\hline
10 & -0.0361037 & -0.0142012 & 0.0259407 & 0.0104886 & 0.00658638
\st\\[1ex]
\hline \hline
$\infty$ &-0.0373037  &-0.014226  &  &
&
\st\\[1ex]
\hline \hline
$conj$ &-0.037483  &-0.0143868  &0.0270638  &0.0105868  &0.00661925
\st\\[1ex]
\hline
\end{tabular}
\end{center}
\caption{The numerical results for the coefficients $r_{n,m}$ for the
tachyon vacuum solution at level $(L,3L)$ approximation,
their  extrapolation to $L=\infty$ via a fit $a + b/L$,
and the conjectured ratios.} \label{t2} \end{table}

The above results show that the vacuum solution gives  
coefficients $r_{m,n}$ for odd $m$, $n$ that are consistent
with the predictions. We now investigate further the universality
of the predictions by examining other solutions of OSFT.

\subsection{Tachyon lump solution}

In this subsection
we present the results for $r_{m,n}$ for the codimension one tachyon
lump solution. Unfortunately due to 
lack of results beyond level (4,8),  we can
carry out the analysis of this and the next subsection only for fields up
to level 4, {\it i.e.} for the coefficients $r_{1,1}$ and $r_{3,1}$.

As in \cite{0005036}, we compactify the direction
transverse to the lump on a circle of radius $R$ so that the momentum in
that direction is quantized in units of $1/R$, and define the level of a
field to be the total $L_0$ eigenvalue of the corresponding state plus one.
Since we are interested in
studying the solution for different values of $R$, in order to carry out
the truncation to a given level, we need to work with different sets of
fields and interaction terms for different values of $R$. We choose, however,
to work with a fixed approximation to the Lagrangian that includes:
\begin{enumerate}
\item All fields up to level 4 and interactions up to level 8 in the zero
momentum sector.

\item Modes 
of the tachyon carrying momentum $\pm n/R$ for $n\le 3$ 
and all
interaction terms involving them.

\item All fields carrying momentum $\pm n/R$ and with level $2+n^2/R^2$
with $n=1$ or 2, and
the interactions among these and the other fields listed above provided
the total level $a+b/R^2$ of all the fields satisfies $a+b/3 <8$.
\end{enumerate}

This choice of  interactions ensures that for $R^2<3$, the
approximation includes all the fields up to level 4 and interactions up to
level 8.
We present the results for $r_{1,1}$ and
$r_{3,1}$ computed with this action for various values of $R$ in the
range $(1, \sqrt{3})$ in table 
\ref{t3}.
The results are again in good agreement with the
predictions
\refb{e21}.\footnote{Although for larger values of
$R$ the ratio $r_{1,1}$ deviates from the expected value $0.407407$
by about 10\%, we note from table \ref{t1} that at level (4,12)
 the vacuum solution ratio $r_{1,1}$ also deviates from the
conjectured value by about 10\%.}
For comparison we have also included the results for
$r_{2,2}$.
The table clearly shows the lack of universality of this coefficient.

\begin{table}
\begin{center}\def\st{\vrule height 3ex width 0ex}
\begin{tabular}{|l|l|l|l|l|l|l|l|} \hline
$R^2$ & 1.05 & 1.1 & 1.5 & 2 & 2.5 & 2.9
\st\\[1ex]
\hline
\hline
$r_{1,1}$ &0.400841   &0.39765  &0.38592  & 0.379426  &0.376123  &
0.374565
\st\\[1ex]
\hline
$r_{3,1}$ &-0.108368  & -0.108068 &-0.107668  & -0.10666  & -0.105063  &
-0.103675
\st\\[1ex]
\hline \hline
$r_{2,2}$ & -0.0312319 & -0.0336458 & -0.0422756 & -0.0471878 & -
0.0497961
&
-0.0510013
\st\\[1ex]
\hline
\end{tabular}
\end{center}
\caption{The numerical results for the coefficients $r_{n,m}$ for the
tachyon lump solution for different values of the radius $R$ of the
compactified direction transverse to the lump.
} \label{t3} \end{table}

\subsection{Marginal deformations}

The set-up here is that of ref.\cite{MARG}. We choose one particular
coordinate tangential to the D-brane which we call $X$, and take the
marginal operator $V$ to be $\p X$. Thus the corresponding state $|V\ra_m$
is given by $\alpha^X_{-1}|0\ra_m$, where $\alpha^X_n$ denote the 
oscillators associated with the field $X$. 
We take the Siegel gauge string  field $|\Psi\ra$ to be of the form:
\be \label{e31}
|\Psi\ra = \lambda \, \alpha^X_{-1} |0\ra_m \otimes c_1|0\ra_g +
|\chi\ra\, ,
\ee
where $|\chi\ra$ satisfies
\be \label{echi}
\la \chi | c_0 ( \alpha^X_{-1} |0\ra_m \otimes c_1|0\ra_g) = 0\, .
\ee
We then 
determine $|\chi\ra$ as a function of $\lambda$ by solving the
components of the
OSFT equations of motion along every state {\it except along
$\alpha^X_{-1}
|0\ra_m \otimes c_1|0\ra_g$.} The general philosophy behind this procedure
is that while we expect the effective potential for $\lambda$ to vanish in
the full string field theory, at any finite level approximation there is a
potential for $\lambda$. Thus the component of the string field equations
along $\alpha^X_{-1} |0\ra_m \otimes c_1|0\ra_g$ will not be satisfied at
finite level approximation for arbitrary $\lambda$. However, since this is
an artifact of level
truncation, we do not insist on satisfying the equations of motion along
this particular direction in the field space.

We work at  level (4,8)  for different values of $\lambda$
and compute the ratios $r_{1,1}$ and $r_{3,1}$ for the solution. The
results are reported in table \ref{t4}. We again find good agreement with
the predictions \refb{e21}. In this context we note that the agreement of
these results with \refb{e21} for small $\lambda$ is automatic since that
is how we
arrived at these predictions in the first place. But the agreement for
finite $\lambda$ provides a non-trivial check on the universality
hypothesis.\footnote{Note that $\lambda\simeq .33$ is the limiting value
beyond which this procedure of obtaining the solution of the string field
equations of motion breaks down\cite{MARG}.}  
We have also displayed in the table the
ratios $r_{2,2}$
computed in this approximation for different values of $\lambda$. The
table clearly shows the lack of universality of these coefficients.

\begin{table}
\begin{center}\def\st{\vrule height 3ex width 0ex}
\begin{tabular}{|l|l|l|l|l|l|} \hline
$\lambda$ & .05 & .1 & .2 & .3  & .32
\st\\[1ex]
\hline
\hline
$r_{1,1}$ & 0.407124 & 0.406234 & 0.401951  & 0.39045 & 0.38568
\st\\[1ex]
\hline
$r_{3,1}$ & -0.109656 &  -0.109408 & -0.108443 & -0.107  & -0.106524
\st\\[1ex]
\hline\hline
$r_{2,2}$ &-0.0262904 &  -0.0270022 & -0.0303961 & -0.0394409  &
-0.0432926
\st\\[1ex]
\hline
\end{tabular}
\end{center}
\caption{The numerical results for the coefficients $r_{n,m}$ for the
solution representing marginal deformation by a Wilson line for different
values of the deformation parameter $\lambda$.
} \label{t4} \end{table}

\sectiono{Possible origin of the universality} \label{sposs}

The universality discussed in the previous section is quite
surprising because given the non-linear nature of the string field
equations of motion, the universal terms
receive contribution from the general non-universal terms. 
This seems
to suggest that the solution of the string field equation satisfies a
set of linear equations which determine the universal part of the
solution. One set of linear constraints was already used to test
the consistency of the universal sector. The universal sector is
$SU(1,1)$ invariant and thus annihilated by the $SU(1,1)$ generators.

At present we do not know of any way to generate a set of linear
equations which completely determine the universal part of the solution,
but we shall now present two sets of linear equations of this type. We
begin with the string field equation:
\be \label{ean1}
Q_B |\Psi\ra + |\Psi * \Psi\ra\, =0 ,
\ee
and apply $c(\pm i)$ on both sides. Since $c$ is an operator of negative
dimension, the action of $c(\pm i)$ on the $*$-product
of two states
vanishes.\footnote{This is true for Fock space states and arises because
the conformal map needed for a midpoint insertion produces
a factor of zero for the case of a negative dimension operator (see, for
example
the related discussion in section 2.1 of \cite{0111129}). We shall proceed
by assuming that this holds for all {\it allowed} string field
configurations. It is
tempting to conjecture that validity of this condition can be taken as the
criterion to determine which string field configurations are allowed.}
This gives
\be \label{ean2}
c(\pm i) Q_B |\Psi\ra = 0\, .
\ee
(Related identities were discussed in ref.\cite{0111087}.)
We note 
in passing that since $c\partial c$ is also of negative
dimension the equation $c\partial c (\pm i) Q_B |\Psi\ra = 0$ must also hold.

Applying $b_0$ on both
sides of \refb{ean2}, and using the results:
\be \label{ean3}
b_0|\Psi\ra = 0, \qquad \{b_0, c(\pm i)\} = \mp i\, , \qquad
\{b_0,Q_B\}=L_0\, ,
\ee
we get
\be \label{ean4}
(\mp i Q_B - c(\pm i) L_0 ) |\Psi\ra = 0\, .
\ee
Taking the sum and differences of these equations, and the expansion
\be \label{ean5}
c(z) = \sum_n c_n z^{-n-1}\, ,
\ee
we get the two linear conditions on $|\Psi\ra$:
\be \label{ean6}
\sum_{m=0}^\infty (-1)^m (c_{2m+1} - c_{-2m-1}) L_0 |\Psi\ra = 0\, ,
\ee
and
\be \label{ean7}
\Big(Q_B - c_0 L_0 - \sum_{m=1}^\infty (-1)^m (c_{-2m} +
c_{2m}) L_0 \Big)|\Psi\ra = 0\, .
\ee

Note in particular that the contribution to the left hand side of
eq.\refb{ean6} along states of the form $c_{-p}c_1|0\ra$ for odd $p$
involves only the components of $|\Psi\ra$
along states of the form $c_{-n} b_{-m} c_1|0\ra$ for odd $m,n$. Since
these coefficients are conjectured to be given by eq.\refb{e12}, one could
ask if they satisfy
eq.\refb{ean6}. Using eqs.\refb{e12}, \refb{ean6} we get,
\be \label{ean8}
(2n+1) \sum_{m=0}^\infty (-1)^m
\widetilde N^{11}_{(2m+1),(2n+1)} = (-1)^{n}\, .
\ee
To establish this equality, we now recall that 
the relevant matrices in the ghost vertex satisfy
\cite{gj2,0111281,0201015}
\be
\label{ean8888}
\wt V^{11}_{nm} = \wt N^{11}_{mn} n \,, \quad \wt M = C \wt V^{11} = - 
E \Bigl( {M\over 1+ 2M}\Bigr)  E^{-1}\,, \quad E_{mn} = \sqrt{m} \delta_{mn} \,, 
\ee
where $M$ is the matrix associated with matter Neumann
coefficients, defined in \cite{0111281}. Equation \refb{ean8} then becomes 
\be \label{ean88}
 \sum_{m=0}^\infty 
\wt V^{11}_{(2n+1),(2m+1)}(-1)^m = -(-1)^{n}\, ,
\ee
which is an eigenvalue equation for a vector $v$:
\be 
\sum_{m=0}^\infty 
 \wt 
V^{11}_{nm}v_m
= - v_n \, \,, \qquad v_{2m+1} = (-1)^m, \qquad v_{2m}=0\, .
\ee
Indeed this equation is satisfied as we now explain. 
Using \refb{ean8888}, we see that the $C$-odd eigenvector $v^-$ of 
$M$ with eigenvalue $-1/3$ satisfies
$\wt V^{11} (E v^-) = - (Ev^-)$. One then readily confirms that
$v = Ev^-$ using the known components of $v^-$ \cite{0111069,0111281}.

A direct proof of \refb{ean8} 
can be given as follows. We have the equation:
\be \label{ean9}
b_0 \big(c(i)+c(-i)\big) \big( ( c_1|0\ra)
* (c_1|0\ra) \big) = 0\, .
\ee
{}From this we can derive eq.\refb{ean8} by expressing
the $*$-product
on the left hand side of eq.\refb{ean9} in terms of the ghost Neumann
coefficients, and then setting to zero the coefficient of $c_{-p}c_1|0\ra$
for odd $p$.
Alternatively we can
argue that since the coefficients $r_{n,m}$ given in
\refb{e12}
explicitly appear in the order $\lambda^2$ solution generated by a
marginal deformation, and since this particular solution also satisfies
the identities \refb{ean6}, \refb{ean7}, eq.\refb{e12} must be consistent
with the
identities \refb{ean6}, \refb{ean7}.

One could also ask if it is possible to check the other identities
following from eqs.\refb{ean6}, \refb{ean7} using the level truncated
solution. Unfortunately the convergence of these relations is not
sufficiently fast to allow us to draw any definite conclusion. In
particular even in the simplest case of $n=0$, the contribution to the
left hand side of eq.\refb{ean8} from terms up to level 14 (e.g. with
$m\le
6$) only gives about 68\% of the expected value of one.  
Going up to level 200 produces about 87\% of the expected answer.

\sectiono{Approximate factorization of $L_0|\Psi_0\ra$} \label{sfactor}

In vacuum string field theory, the solutions of classical equations of
motion factorize into a product of a state in the matter state space and
another state in the ghost state space. Since in OSFT the kinetic operator
$Q_B$ receives contribution from both the ghost and the matter sector, we
do not expect the solutions to have such simple factorization property.
While the vacuum solution
$|\Psi_0\ra$ is far from factorized,  $L_0|\Psi_0\ra$ seems to possess 
some
approximate factorization property which we shall demonstrate now.

We begin by giving $L_0 |\Psi_0 \ra$, calculated at level (10,30),
up to level 6:
\ben\label{factor2}
L_0 |\Psi_0\ra &=&  -0.54626 \, \Big[1 -  \,
0.10478 L^m_{-2}  +0.39721 \,b_{-1} c_{-1}
 +0.02798 \,L^m_{-4}  \\
&&      +0.00354 \,L^m_{-2} L^m_{-2}
-\underbar{0.04322}              \,L^m_{-2} b_{-1} c_{-1}
-0.32011                \,b_{-1} c_{-3}  \nonumber \\
&& -0.19211             \,b_{-2} c_{-2}
- 0.10670       \,b_{-3} c_{-1}   -0.01327\,L^m_{-6}
\nonumber \\
&& -0.00282 \,L^m_{-4} L^m_{-2}   - 0.00006 \,  L^m_{-3} L^m_{-3}
+0.00007 \,L^m_{-2} L^m_{-2} L^m_{-2}
\nonumber \\
&&+\underbar{0.01110}      \,L^m_{-4} b_{-1} c_{-1}
        +\underbar{0.00173}\,L^m_{-2} L^m_{-2}     b_{-1} c_{-1}
+\underbar{0.00068}\,L^m_{-3}      b_{-1} c_{-2}
\nonumber \\
&& +\underbar{0.00034}     \,L^m_{-3}b_{-2} c_{-1}
      +\underbar{0.03217}          \,L^m_{-2}      b_{-1} c_{-3}
+\underbar{0.02708}        \,L^m_{-2}      b_{-2} c_{-2}
\nonumber \\
&& +\underbar{0.01072}\,L^m_{-2}   b_{-3} c_{-1}  +0.27924 \,b_{-1} c_{-5}
+0.17495        \,b_{-2} c_{-4}
\nonumber \\
&& +0.10496     \,b_{-3} c_{-3} -0.07393
\,b_{-2} b_{-1} c_{-1} c_{-2}
+0.08747 \,             b_{-4} c_{-2}
\nonumber \\
&& +0.05584\,   b_{-5} c_{-1}     \, \Big] \, \ta \,.  
\nonumber \een
In the above, we have underlined the terms
that involve both matter and ghost operators. 
If $L_0|\Psi_0\rangle$ had a factorized form, then we could
determine the state by  simply looking at the pure matter and pure ghost
excitations in
$L_0|\Psi_0\ra$ and then taking their direct product. Calling
$(L_0|\Psi_0\ra)_0$ the state assembled in this way from the above  
equation, we have:
\ben \label{efactor3}
(L_0|\Psi_0\ra)_0 &=& -0.54626 \, \Big( 1 +0.39721 \,b_{-1} c_{-1}
 -0.32011               \,b_{-1} c_{-3} \\
&& -0.19211             \,b_{-2} c_{-2}   - 0.10670     \,b_{-3}
c_{-1}+\cdots   \Big) \cdot
\nonumber \\
  && \Big(1    
        - 0.10478 \,L^m_{-2}
 +0.02798 \,L^m_{-4}  +0.00354 \,L^m_{-2} L^m_{-2} +\cdots \Big) \ta
\nonumber\\
&=&      -0.54626 \, \Big[1 -   \,
0.10478 L^m_{-2}  +0.39721 \,b_{-1} c_{-1}
 +0.02798 \,L^m_{-4} \nonumber \\
&&      +0.00354 \,L^m_{-2} L^m_{-2}
 -\underbar{0.04162}      \,L^m_{-2} b_{-1} c_{-1}
-0.32011                \,b_{-1} c_{-3}  \nonumber \\
&& -0.19211             \,b_{-2} c_{-2}
- 0.10670       \,b_{-3} c_{-1}   -0.01327\,L^m_{-6}
\nonumber \\
&& -0.00282 \,L^m_{-4} L^m_{-2}   - 0.00006\, L^m_{-3} L^m_{-3} 
0.00006
+0.00007 \,L^m_{-2} L^m_{-2} L^m_{-2}
\nonumber \\
&&+\underbar{0.01111} \,L^m_{-4} b_{-1} c_{-1}
        + \underbar{0.00141}      \,L^m_{-2} L^m_{-2}      b_{-1} c_{-1}
+ \underbar{0.0}    \,L^m_{-3}       b_{-1} c_{-2}
\nonumber \\
&& +    \underbar{0.0}       \,L^m_{-3}b_{-2} c_{-1}
   +   \underbar{0.03354}          \,L^m_{-2}      b_{-1} c_{-3}
+\underbar{0.02013}        \,L^m_{-2}      b_{-2} c_{-2}
\nonumber \\
&& +\underbar{0.01118}          \,L^m_{-2}         b_{-3} c_{-1}  +0.27924
\,b_{-1} c_{-5}
+0.17495        \,b_{-2} c_{-4}
\nonumber \\
&& +0.10496     \,b_{-3} c_{-3} -0.07393
\,b_{-2} b_{-1} c_{-1} c_{-2}
+0.08747 \,             b_{-4} c_{-2}
\nonumber \\
&& +0.05584\,   b_{-5} c_{-1}     \, \Big] \, \ta \,.   
\nonumber
\een
We now compare the expressions on the right hand sides of eq.\refb{factor2}
and \refb{efactor3}. 
By construction, the pure matter and pure ghost terms are identical.
The underlined terms, however, test the factorization property since
they arise from products. 
Some terms are remarkably accurate (like the $L_{-4}^m b_{-1} c_{-1}$, with
error of one part in a thousand), several are within about 5\%, and a couple
exceed 20\% error.  Experiments with lower level results  
indicate that the coefficients do not always approach each other as we
increase the level. If factorization held exactly,
terms  as $L_{-3} b_{-2} c_{-1}$ would have to vanish since because of
twist property the string field cannot have expectation values for the
separate matter and ghost parts. Indeed, the coefficients of such terms
in $L_0|\Psi_0\ra$ are small, but they do not seem to go 
to zero as the level is increased.
Overall, despite striking patterns, we conclude that $L_0 |\Psi_0\rangle$ 
is at best approximately factorized.  

Since the star product of factorized fields is factorized, 
an approximate factorization for $L_0 |\Psi_0\rangle$ would arise  
if the dominant  contribution to $b_0|\Psi_0*\Psi_0\ra$  
(and hence, $L_0|\Psi_0\ra$) came from a part
of $|\Psi_0\ra$ that is factorized. To see if this is the origin of
factorization, we have taken the level 10 expression for $|\Psi_0\ra$,
identified its pure matter and pure ghost excitations, and defined a new
configuration $|\Phi_0\ra$ by taking the direct product of these two
factors. The result for $|\Phi_0*\Phi_0\ra$ is given below:

\ben \label{efactor4}
-b_0|\Phi_0*\Phi_0\ra & =&
  -0.45634 \, \Big[1 -
0.06884 L^m_{-2}  +   0.39105  \,b_{-1} c_{-1}
 +  0.02722 \,L^m_{-4}  \\
&&     -0.00170 \,L^m_{-2} L^m_{-2}
-\underbar{  0.02707 }              \,L^m_{-2} b_{-1} c_{-1}
-0.32889     \,b_{-1} c_{-3}  \nonumber \\
&& -0.15360             \,b_{-2} c_{-2}
 -0.10963     \,b_{-3} c_{-1}    -0.01305 \,L^m_{-6}
\nonumber \\
&& -0.00161 \,L^m_{-4} L^m_{-2}   - 0.00014 \,  L^m_{-3} L^m_{-3}
+0.00042 \,L^m_{-2} L^m_{-2} L^m_{-2}
\nonumber \\
&&+\underbar{ 0.01067}      \,L^m_{-4} b_{-1} c_{-1}
        -\underbar{0.00064}\,L^m_{-2} L^m_{-2}     b_{-1} c_{-1}
+\underbar{0.0}\,L^m_{-3}      b_{-1} c_{-2}
\nonumber \\
&& +\underbar{0.0}     \,L^m_{-3}b_{-2} c_{-1}
      +\underbar{  0.02277}          \,L^m_{-2}      b_{-1}
c_{-3}
+\underbar{0.01059}        \,L^m_{-2}      b_{-2} c_{-2}
\nonumber \\
&& +\underbar{0.00759}\,L^m_{-2}   b_{-3} c_{-1}  +0.28947 \,b_{-1}
c_{-5}
+0.14175        \,b_{-2} c_{-4}
\nonumber \\
&& +0.10522     \,b_{-3} c_{-3} -0.05942
\,b_{-2} b_{-1} c_{-1} c_{-2}
+0.07087 \,             b_{-4} c_{-2}\nonumber \\
&& +0.05789\,   b_{-5} c_{-1}     \, \Big] \, \ta \,.
\nonumber
\een
Comparing eqs.\refb{factor2}-\refb{efactor4} we see that \refb{efactor3}
is 
closer to \refb{factor2}
than \refb{efactor4} is to
\refb{factor2}. Thus the explanation for the approximate factorization
{\it does not} completely 
 lie in the fact that the factorized part of $|\Psi_0\ra$ gives the
dominant contribution to $|\Psi_0*\Psi_0\ra$. Somehow, products of 
factorized
terms
times mixed terms, and mixed terms times mixed terms do give substantial
factorized contributions.

\sectiono{A surprising coincidence in the matter sector} \label{ssurp}

In  previous sections we discussed a possible universality of  the ghost 
part of
the OSFT field equations. This analysis shows that at least some part of
the solution involving ghost excitations may not depend on  
which particular background the solution describes. In this
section we shall discuss a different kind of universality in the matter
sector where we shall show that two different field equations which differ
from each other in their ghost structure have
solutions whose pure matter
parts appear to be quite close to each other. 

The OSFT field equation in the Siegel gauge takes the form:
\be \label{e51}
L_0 |\Psi\ra = -b_0 |\Psi *\Psi\ra\, ,
\ee
where $|\Psi\ra$ is a ghost number 1 field equation. The solution of this
equation representing the tachyon vacuum solution is a linear combination
of states, obtained by 
matter Virasoro generators
$L^{(m)}_{-k}$
and ghost oscillators $b_{-m}$, $c_{-n}$ acting on
$c_1|0\ra$\cite{9911116}. Let us
focus on the part of the solution involving pure matter excitations. If we
denote by $\OO_s(L^{(m)})$ some combination of matter Virasoro generators,
then the pure matter part of the solution takes the form:
\be \label{e53}
\alpha_0 \Big( c_1|0\ra + \sum_s \beta_s \OO_s(L^{(m)}) c_1|0\ra\Big)\, ,
\ee
where $\alpha_0$, $\beta_s$ are constants. We adopt the convention that
$\beta_{-m_1, \ldots -m_r}$ will denote the coefficient of $L^{(m)}_{-m_1}
\ldots
L^{(m)}_{-m_r} c_1|0\ra$,
with $m_1 \geq m_2 \geq \dots \geq m_r$.

Since $L_0$ acting on
the ghost vacuum $c_1|0\ra$ has eigenvalue $-1$, a closely related
equation for a ghost number zero field $|\Phi\ra$ is:
\be \label{e52}
(L_0-1)|\Phi\ra = - |\Phi * \Phi\ra\, .
\ee
We can look for a solution of this equation of the form:
\be \label{e54}
|\Phi_0\ra = |0\ra + \sum_s \gamma_s \OO_s(L) |0\ra\, ,
\ee
where $\gamma_s$ are constants and $L$ denotes the total Virasoro
generators of the matter-ghost system. We again use the convention that
$\gamma_{-m_1, \ldots -m_r}$ denotes the coefficient of $L_{-m_1}\ldots
L_{-m_r}|0\ra$. Note that since the Virasoro
generators $L_m$ have vanishing central charge, the coefficient of $|0\ra$
in eq.\refb{e54} is unity. From eq.\refb{e54} it follows that part of
$|\Phi_0\ra$ involving pure matter excitations is given by:
\be \label{e55}
|0\ra + \sum_s \gamma_s \OO_s(L^{(m)}) |0\ra\, .
\ee

Examining the numerical solutions of eqs.\refb{e51} and \refb{e52} using
level truncation, we find
the following correspondence between the coefficients $\beta_s$
and $\gamma_s$:
\be \label{e56}
\beta_s \simeq \gamma_s\, .
\ee
The results are presented in appendix \ref{sa}.
For most of the coefficients the correspondence seems to be valid to
within a few percent at level 10. 
There are some exceptions, {\it e.g.}
the coefficients of  
$(L_{-2})^n$, with $n\geq 3$. However we note from the tables in
appendix \ref{sa}
that at a given level, the coefficients of such terms are order of
magnitude smaller than the leading terms at this level, {\it e.g.} the
coefficient of $L_{-2n}$. Indeed we see from these tables that the error
in the coefficients of $(L_{-2})^n$ is
of the same order of magnitude as
the error in the coefficients of $L_{-2n}$.

\sectiono{Concluding remarks}

One lesson we have learned in the present investigation
is that ``quasi-patterns" seem to exist -- these are
remarkable coincidences having  a theoretical ring to
them that appear to hold closely  in the
level expansion but do not truly hold exactly.
A pattern may be only a  quasi-pattern
when successive level calculations do not appear to improve
systematically the accuracy.  Perhaps the first example
of such quasi-pattern was the zero norm property of
Ref.~\cite{0009105}, the accuracy of which
stops improving at level eight.

We have provided evidence that certain ratios of
expectation values in the purely ghost sector of all
(known) solutions of OSFT take fixed values given
by simple expressions in terms of Neumann coefficients.
While we have no full understanding of the theoretical
meaning of this observation, we showed that linear constraints
on the string field give partial consistency checks
and thus some analytic evidence for the proposal.
The numerical evidence also seems to improve with level. All in
all we feel that there is some strong but not overwhelming
evidence for this to be an exact pattern.

For the case of the correspondence between ghost number
zero and ghost number one field equations, the near
equality of expectation values is
quite striking, but the pattern is somewhat irregular
and lacking an analytic understanding for how
this correspondence could arise, 
it seems premature
to propose a strict equality.
In particular, for some coefficients ({\it e.g.} of $L_{-4}$) where the
results are fairly stable in level expansion, the correspondence still has
an error of about 2\%.
Higher level data could
help us reach a conclusion, but this could be a quasi-pattern.

Finally, the factorization of $L_0\Psi_0$ for the tachyon
vacuum solution would seem to be a quasi-pattern.  Given
the theoretical significance of factorized states
(such as surface states) the factorization pattern would
be remarkable if exact, but the evidence again is that
some small errors do not seem to get smaller as we proceed at
higher levels.

It is hoped that the observations and patterns noted
in this work will help guide in the search for
exact analytic solutions of OSFT.

\bigskip
\noindent
{\bf Acknowledgements.}  
B.Z. would like to thank the Rutgers University
Physics Department for hospitality during the period
when this research was started.
The work of L.R. was supported in part
by Princeton University
``Dicke Fellowship'' and by NSF grant 9802484.
The  research of A.S. was supported in part by a grant 
from the Eberly College
of Science of the Penn State University.
The work of  B.Z. was supported in part
by DOE contract \#DE-FC02-94ER40818.

\appendix

\sectiono{Comparison of $\beta_s$ and $\gamma_s$} \label{sa}

In this appendix we shall give the results for the coefficients $\beta_s$
and $\gamma_s$ at various levels of approximation. The $L=\infty$ answer
given in each table is obtained using an extrapolation in $L$ with a
function $a + b/L$.

\begin{center}\def\st{\vrule height 3ex width 0ex}
\begin{tabular}{|l|l|l|l|l|l|l|l|l|} \hline
$L$ & $\beta_{-2}$ & $\gamma_{-2}$ && $\beta_{-4}$  &
$\gamma_{-4}$ && $\beta_{-2,-2}$ & $\gamma_{-2,-2}$
\st\\[1ex]
\hline
\hline
4 & 0.103799  & 0.106693  && -0.009338  & -0.00941869  &&-0.00107467  &
-0.00119004 \st\\[1ex]
\hline
6 &0.104289  &0.106822  &&-0.00932407  & -0.00945266  &&-0.00113529  &
-0.00121053 \st\\[1ex]
\hline
8 & 0.104586  &0.106864  &&-0.0093259  & -0.00946405  &&-0.00116427  &
-0.00121627 \st\\[1ex]
\hline
10 &0.104787  &0.106882  &&-0.00932777  & -0.00946906  &&-0.00118174
&-0.00121856 \st\\[1ex]
\hline \hline
$\infty$ &0.10541 &0.107021 &&-0.00931707  &-0.009506  &&-0.00125357
 &-0.00124007
\st\\[1ex]\hline
\end{tabular}
\end{center}

\begin{center}\def\st{\vrule height 3ex width 0ex}
\begin{tabular}{|l|l|l|l|l|l|} \hline
$L$ & $\beta_{-6}$ & $\gamma_{-6}$ && $\beta_{-4,-2}$  &
$\gamma_{-4,-2}$
\st\\[1ex]
\hline
\hline
6 & 0.00266208  & 0.00267922  &&0.00055773  & 0.000588318  \st\\[1ex]
\hline
8 &0.00265609  &0.00268466  &&0.000562876  & 0.000591376  \st\\[1ex]
\hline
10 & 0.00265563  &0.00268703  &&0.000565538  &0.000592452  \st\\[1ex]
\hline\hline
$\infty$ &0.00264468  &0.00269687  &&0.000577424  &0.000598963  \st\\[1ex]
\hline
\end{tabular}
\end{center}

\begin{center}\def\st{\vrule height 3ex width 0ex}
\begin{tabular}{|l|l|l|l|l|l|} \hline
$L$ & $\beta_{-3,-3}$ & $\gamma_{-3,-3}$ && $\beta_{-2,-2,-2}$  &
$\gamma_{-2,-2,-2}$
\st\\[1ex]
\hline
\hline
6 &1.1188$\times 10^{-5}$  &1.14982$\times 10^{-5}$  &&-1.80854$\times
10^{-5}$  & -1.81025$\times 10^{-5}$  \st\\[1ex]
\hline
8 &1.27096$\times 10^{-5}$  &1.30298$\times 10^{-5}$  &&-1.56105$\times
10^{-5}$  & -1.74459$\times 10^{-5}$  \st\\[1ex]
\hline
10 &1.3302$\times 10^{-5}$  &1.35892$\times 10^{-5}$  &&-1.42765$\times
10^{-5}$  & -1.72127$\times 10^{-5}$ \st\\[1ex]
\hline\hline
$\infty$ &1.66038$\times 10^{-5}$  &1.68725$\times 10^{-5}$
&&-0.850154$\times 10^{-5}$  &   -1.58124$\times 10^{-5}$
\st\\[1ex]
\hline
\end{tabular}
\end{center}

\begin{center}\def\st{\vrule height 3ex width 0ex}
\begin{tabular}{|l|l|l|l|l|l|} \hline
$L$ & $\beta_{-8}$ & $\gamma_{-8}$ && $\beta_{-6,-2}$  &
$\gamma_{-6,-2}$
\st\\[1ex]
\hline
\hline
8 &-0.00104572  &-0.00104938  &&-0.000190741  & -0.000199537  \st\\[1ex]
\hline
10 &-0.00104339  &-0.00105092  && -0.000191915  & -0.000200248 \st\\[1ex]
\hline\hline
$\infty$ &-0.00103407  &-0.00105708  &&-0.000196611  &-0.000203092
\st\\[1ex]
\hline
\end{tabular}
\end{center}

\begin{center}\def\st{\vrule height 3ex width 0ex}
\begin{tabular}{|l|l|l|l|l|l|} \hline
$L$ & $\beta_{-5,-3}$ & $\gamma_{-5,-3}$ && $\beta_{-4,-4}$  &
$\gamma_{-4,-4}$
\st\\[1ex]
\hline
\hline
8 &-3.30137$\times 10^{-6}$  &-3.21237$\times 10^{-6}$  &&-5.50356$\times
10^{-5}$  &  -5.67779$\times 10^{-5}$ \st\\[1ex]
\hline
10 &-3.76634$\times 10^{-6}$  &-3.73539$\times 10^{-6}$  &&
-5.51104$\times 10^{-5}$ &-5.69378$\times 10^{-5}$  \st\\[1ex]
\hline\hline
$\infty$ &-5.62622$\times 10^{-6}$  &-5.82747$\times 10^{-6}$
&&-5.54096$\times 10^{-5}$  &  -5.75774$\times 10^{-5}$
\st\\[1ex]
\hline
\end{tabular}\end{center}

\begin{center}\def\st{\vrule height 3ex width 0ex}
\begin{tabular}{|l|l|l|l|l|l|} \hline
$L$ & $\beta_{-4,-2,-2}$ & $\gamma_{-4,-2,-2}$ && $\beta_{-3,-3,-2}$  &
$\gamma_{-3,-3,-2}$
\st\\[1ex]
\hline
\hline
8 &-1.09361$\times 10^{-5}$  &-1.25777$\times 10^{-5}$  &&-1.35659$\times
10^{-6}$  & -1.49837$\times 10^{-6}$  \st\\[1ex]
\hline
10 &-1.14282$\times 10^{-5}$  &-1.27416$\times 10^{-5}$  &&-1.50681$\times
10^{-6}$  & -1.62021$\times 10^{-6}$ \st\\[1ex]
\hline\hline
$\infty$ &-1.33966$\times 10^{-5}$  &-1.33972$\times 10^{-5}$
&&-2.10769$\times 10^{-6}$  &  -2.10757$\times 10^{-6}$
\st\\[1ex]
\hline
\end{tabular}
\end{center}

\begin{center}\def\st{\vrule height 3ex width 0ex}
\begin{tabular}{|l|l|l|} \hline
$L$ & $\beta_{-2,-2,-2,-2}$ & $\gamma_{-2,-2,-2,-2}$
\st\\[1ex]
\hline
\hline
8 &1.81597$\times 10^{-6}$  & 2.08577$\times 10^{-6}$ \st\\[1ex]
\hline
10 &1.73594$\times 10^{-6}$ &2.06738$\times 10^{-6}$  \st\\[1ex]
\hline\hline
$\infty$ &1.41582$\times 10^{-6}$  &1.99382$\times 10^{-6}$ \st\\[1ex]
\hline
\end{tabular}
\end{center}

\begin{center}\def\st{\vrule height 3ex width 0ex}
\begin{tabular}{|l|l|l|l|l|l|} \hline
$L$ & $\beta_{-10}$ & $\gamma_{-10}$ && $\beta_{-8,-2}$  &
$\gamma_{-8,-2}$
\st\\[1ex]
\hline
\hline
10 & 0.000533168 &0.00053466  &&8.26156$\times 10^{-5}$  & 8.57127$\times
10^{-5}$ \st\\[1ex]
\hline
\end{tabular}
\end{center}

\begin{center}\def\st{\vrule height 3ex width 0ex}
\begin{tabular}{|l|l|l|l|l|l|} \hline
$L$ & $\beta_{-7,-3}$ & $\gamma_{-7,-3}$ && $\beta_{-6,-4}$  &
$\gamma_{-6,-4}$
\st\\[1ex]
\hline
\hline
10 & 7.80062$\times 10^{-7}$ &7.17496$\times 10^{-7}$  &&4.07104$\times
10^{-5}$  & 4.18855$\times 10^{-5}$ \st\\[1ex]
\hline
\end{tabular}
\end{center}

\begin{center}\def\st{\vrule height 3ex width 0ex}
\begin{tabular}{|l|l|l|l|l|l|} \hline
$L$ & $\beta_{-5,-5}$ & $\gamma_{-5,-5}$ && $\beta_{-6,-2,-2}$  &
$\gamma_{-6,-2,-2}$
\st\\[1ex]
\hline
\hline
10 & 3.14321$\times 10^{-7}$  &2.84212$\times 10^{-7}$  &&4.23563$\times
10^{-6}$  &4.70925$\times 10^{-6}$  \st\\[1ex]
\hline
\end{tabular}
\end{center}

\begin{center}\def\st{\vrule height 3ex width 0ex}
\begin{tabular}{|l|l|l|l|l|l|} \hline
$L$ & $\beta_{-5,-3,-2}$ & $\gamma_{-5,-3,-2}$ && $\beta_{-4,-4,-2}$  &
$\gamma_{-4,-4,-2}$
\st\\[1ex]
\hline
\hline
10 &4.48058$\times 10^{-7}$  &4.83439$\times 10^{-7}$  &&4.097$\times
10^{-6}$  &4.44877$\times 10^{-6}$  \st\\[1ex]
\hline
\end{tabular}
\end{center}

\begin{center}\def\st{\vrule height 3ex width 0ex}
\begin{tabular}{|l|l|l|l|l|l|} \hline
$L$ & $\beta_{-4,-3,-3}$ & $\gamma_{-4,-3,-3}$ && $\beta_{-4,-2,-2,-2}$  &
$\gamma_{-4,-2,-2,-2}$
\st\\[1ex]
\hline
\hline
10 & 1.89101$\times 10^{-7}$  & 1.96492$\times 10^{-7}$ && -4.13579$\times
10^{-7}$  &-4.29091$\times 10^{-7}$  \st\\[1ex]
\hline
\end{tabular}
\end{center}

\begin{center}\def\st{\vrule height 3ex width 0ex}
\begin{tabular}{|l|l|l|l|l|l|} \hline
$L$ & $\beta_{-3,-3,-2,-2}$ & $\gamma_{-3,-3,-2,-2}$ &&
$\beta_{-2,-2,-2,-2,-2}$  &
$\gamma_{-2,-2,-2,-2,-2}$
\st\\[1ex]
\hline
\hline
10 &8.47716$\times 10^{-8}$  &9.94688$\times 10^{-8}$  &&-6.62409$\times
10^{-8}$  & -8.43162$\times 10^{-8}$ \st\\[1ex]
\hline
\end{tabular}
\end{center}

\end{document}